\documentstyle[preprint,aps]{revtex}

\begin{document}

\draft
\preprint{Submitted to Appl.\ Phys.\ Lett.}

\title{
Charge sensitivity of superconducting single-electron
transistor}

\author{Alexander N. Korotkov}
\address{
Department of Physics, State University of New York,
Stony Brook, NY 11794-3800  \\
and\\
Nuclear Physics Institute, Moscow State University, Moscow 119899 Russia}

\date{\today}

\maketitle

\begin{abstract}
        It is shown that the noise-limited charge sensitivity of 
a single-electron transistor using superconductors (of either $SISIS$ 
or $NISIN$ type) operating near the threshold of quasiparticle 
tunneling, can be considerably higher than that
of a similar transistor made of normal metals or semiconductors.
        The reason is that the superconducting energy gap, in contrast 
to the Coulomb blockade, is not smeared by the finite temperature.
We discuss also the increase of the maximum operation temperature
due to superconductivity and a new peak-like feature on the 
$I-V$ curve of $SISIS$ structures.

\end{abstract}

\narrowtext

\vspace{1ex}

        Electron transport in the systems of small-capacitance tunnel
junctions shows a variety of single-electron effects \cite{Av-Likh}.
The simplest and most thoroughly studied circuit revealing these effects
is the so-called Single Electron Transistor \cite{Likh-87}
(SET) which consists of two tunnel junctions 
connected in series. 
At low temperatures ($T \ll e^2/C_\Sigma$, $C_\Sigma =C_1+C_2$ where
$C_1$ and $C_2$ are the junction capacitances) the current through 
this structure depends on the background 
charge $Q_0$ of the central electrode (the dependence is periodical 
with a period
equal to the electron charge $e$). Hence, controlling $Q_0$ (for example, 
by a capacitive gate)  
it is possible to control the current $I$ through the circuit.
	The possibility to use the SET as a highly-sensitive electrometer 
has been confirmed in numerous experiments.

        The most developed technology of the SET fabrication uses the 
overlapping narrow aluminum
films with a typical junction capacitance about few times $10^{-16}$ F
(see, e.g.\ Refs.\ \cite{Lafarge,Tinkham-SET,Mooij}).
Consequently, the operation temperature is typically less than
1 K, and the electrodes are in the superconducting state unless
the superconductivity is intentionally suppressed by the magnetic field. 
It has been noticed \cite{Lafarge,Tinkham-SET,Krech-sc} that the 
superconductivity of electrodes improves the performance of the SET 
(operating near the threshold of quasiparticle
tunneling) as an electrometer  in comparison with the 
normal-state operation. However, we are not aware of 
quantitative theoretical analysis of this issue, which will be the 
subject of the present paper.

         There are two major characteristics of the SET operation 
as an electrometer. The first one is the amplitude
of the output signal modulation for $Q_0$ variations larger than $e$. 
It was found experimentally
\cite{Tinkham-SET} that the use of superconducting electrodes
increases the modulation amplitude of current $I$ (for fixed
bias voltage $V$), especially at temperatures comparable
to $e^2/C_\Sigma$, thus increasing the maximum temperature. The 
theoretical results of the present paper confirm this statement for 
both $NISIN$ and $SISIS$ structures.

        The other, even more important characteristic of the SET operation
is the noise-limited sensitivity (ability to detect 
variations of $Q_0$ much smaller than $e$).
The best achieved sensitivity so far (by the normal state SET) 
is  $7\times 10^{-5} e/\sqrt{Hz}$ at $10 Hz$ \cite{Mooij}.
In the present-day technology this figure is limited by 1/f noise
which is most likely caused
by random trapping-escape processes in nearby impurities.
So, in some sense, the sensitivity does
not depend much on the parameters of the SET, but rather on
the purity of the sample.  It is unlikely that superconductivity
of electrodes can significantly affect these processes. 
Hence, the present-day sensitivities of superconducting and normal SETs
with similar parameters should not differ much for reasonably low 
temperatures when both SETs show sufficient modulation amplitude.

	However, with the technology improvement
one can expect the reduction of the  noise due to impurities. Then
the charge sensitivity of the SET would achieve the limit
determined by the intrinsic noise \cite{SQUID,Theses} of the device 
caused by random electron jumps through tunnel junctions
(this ``white'' noise has been recently
measured in experiment \cite{Birk}). Though the theory of the 
``classical'' thermal/shot intrinsic noise of the SET 
is applicable to the general case of one-particle
tunneling (normal metals, semiconductors, quasiparticle current in
superconductors, etc.), most numerical results in Refs. 
\cite{SQUID} and \cite{Theses} as well as in a number
of subsequent papers on this subject (see, e.g.\ Refs. 
\cite{Hershfield,Chen,Galperin,Krech}) were obtained only for
SETs made of normal metals. (Recently some generalization
was done \cite{Galperin-SC} to include the possibility of two-particle
tunneling which can be important in the superconducting case. Let us also
mention Ref.\ \cite{Krech-sc} in which the noise in $NISIN$ SET was
briefly considered.) 

        In the present paper we apply the theory of Refs.\ 
\cite{SQUID} and \cite{Theses} to the cases of capacitively
coupled superconducting $SISIS$ and $NISIN$ SETs (the analysis of
a resistively coupled SET can be done in a similar way - see Ref.\
\cite{SQUID}). We show that the noise-limited sensitivity of a
SET-electrometer can be considerably improved by the use of superconducting
electrodes. 

        We consider only the quasiparticle tunneling, neglecting the 
Josephson current, resonant tunneling of 
Cooper pairs,  Andreev reflection, and cotunneling.
This assumption is appropriate when the Josephson coupling is 
negligible and the normal state resistances $R_1$ and $R_2$ of tunnel
junctions are well above the resistance quantum $R_Q=\pi\hbar/2e^2$.
	We use the ``orthodox'' theory \cite{Av-Likh,Likh-87}
of the SET and the BCS theory \cite{Tinkham}
for the calculation of the tunneling rates.

        Figure 1  shows the $I-V$ curves 
at different temperatures for (a) the normal metal $NININ$ case, 
(b) $NISIN$ case (which is equivalent to $SINIS$ case), and
(c)--(d) $SISIS$ case. SETs  with $C_1=C_2$ and  
$R_1=R_2=R_\Sigma /2$ are chosen, and 
we neglect the gate capacitance $C_g$ because it can always be formally
distributed between $C_1$ and $C_2$ (see, e.g., Ref.\ \cite{RC-SET}).
Three curves in each set represent $Q_0=0$, $e/4$, and $e/2$, respectively.
Temperature increase decreases the superconducting energy gap
$\Delta (T)$ (which is assumed to be equal in all S-electrodes) 
leading to the noticeable shift to the left of the 
positions of the current jumps in Figs. 1(c) and 1(d). The pure BCS 
theory would lead to
the abrupt jumps of the current in $SISIS$ case. To take
into account the unavoidable smoothing of the jumps in reality,
we assume additionally the inhomogeneous broadening of $\Delta(0)$ with 
Gaussian distribution characterized by the dispersion
$w_0$. This phenomenological parameter is chosen as $w_0=0.05\Delta(0)$
in Figs. 1(c) and 1(d) (for finite temperatures $w(T)=w_0[\Delta (T) /
\Delta(0) -(T/\Delta (0)) (d \Delta (T)/dT)]$ was used).

        One can see that in the normal metal case the current $I$ can
be considerably modulated ($I_{max}/I_{min} \agt 2$) by $Q_0$ 
($V$ is fixed) only at
$T\alt 0.15 e^2/C_\Sigma$, while at $T=0.3 e^2/C_\Sigma$ the modulation
is already negligible, $(I_{max}-I_{min})/I_{max} \simeq 5\%$. Notice that
the maximum relative modulation is achieved at small voltages and does
not depend on ratios $C_1/C_2$ and $R_1/R_2$.

        $NISIN$ transistor with $\Delta(0)=0.5 e^2/C_\Sigma$ 
shows considerable modulation crudely up to $T\approx
0.2 e^2/C_\Sigma$, while $SISIS$ transistors with $\Delta(0)=0.5 
e^2/C_\Sigma$ and $\Delta(0)=2.0 e^2/C_\Sigma$ operate well almost up 
to the critical temperature $T_c$ ($T_c/
(e^2/C_\Sigma)=0.28$ and 1.14 respectively). 
        The case $\Delta(0)=0.5 
e^2/C_\Sigma$ corresponds to the typical present-day
experimental situation with aluminum junctions and $C_\Sigma\approx 0.4$ fF
(see, e.g.\ Ref.\ \cite{Tinkham-SET}).
Comparison of Figs.\ 1(c) and 1(d) shows that the increase of $\Delta(0)$
provides further improvement of the transistor performance at high
temperatures. Using Fig.\ 1(d) one can predict the operation of the
niobium-based SET with $C_\Sigma \approx 0.2$ fF (current 
state-of-the-art for aluminum junctions) at temperatures up to 7 K.

        Superconductivity improves the SET performance at 
relatively high temperatures because, in contrast to the Coulomb 
blockade, the superconducting energy gap is not smeared by the 
finite temperature. In the normal metal case the $I-V$ curve has a cusp
at the Coulomb blockade threshold $V_t=\min_{i,n}\{V_{i,n} \mid V_{i,n}>0\}$,
where
\begin{equation}
        V_{i,n} =\frac{e}{C_i} \left( \frac{1}{2} +(-1)^i 
(n+\frac{Q_0}{e}) \right) \, ,
\label{Vin}\end{equation}
and this cusp is rounded within the voltage interval proportional to the 
temperature. In $SISIS$ case the
jump of the $I-V$ curve at $V_t$ which is shifted due to the energy gap, 
        $V_t=\min_{i,n} \{V_{i,n}+2\Delta (T) C_\Sigma /eC_i 
\mid V_t>4\Delta (T) \}$,
        remains sharp even at $T\sim \Delta(T)$, and the subthreshold 
current increase is only proportional to $\mbox{exp}(-T/\Delta(T))$.
This explains why $SISIS$ transistor shows considerable dependence
on $Q_0$ for the temperatures almost up to $T_c$ even if $T \agt 
e^2/C_\Sigma$. In $NISIN$ case the I-V curve
in the vicinity of $V_t=\min_{i,n} \{V_{i,n}+\Delta (T) C_\Sigma /eC_i 
\mid V_t>2\Delta (T) \}$ is rounded by the finite temperature, 
that makes $NISIN$ transistor
worse than $SISIS$ transistor, however, it is still better than
usual $NININ$ transistor.

        Let us briefly discuss the origin of small peaks of the current at 
moderate temperatures visible in Figs. 1(c) and 1(d) ($SISIS$ case) at
voltages close to the middle of the subthreshold region. The position 
of a peak satisfy Eq.\ (\ref{Vin})
        and corresponds to zero energy gain $W$ for a particular
tunneling process (hence, it coincides with the position of one of the
$I-V$ cusps in the corresponding $NININ$ SET).
In this case the singularities in the 
density of states of two electrodes match, leading to increase of
tunneling of thermally excited quasiparticles. Hence, the origin
of peaks is similar to that of well-known peaks \cite{Tinkham} at
$V=(\Delta_1(T)-\Delta_2(T))/e$ in the single junction made of
superconductors with different gaps $\Delta_1(T)$ and $\Delta_2(T)$. 
In our case energy gaps are the same but the Coulomb blockade provides 
the relative shift of the  singularities in the density of states. 
	The analysis of the master equation \cite{Av-Likh,Likh-87}
shows that the singularity-matching peak can be significantly high 
only within the voltage range $2\Delta (T) <V<
2\Delta (T) +e/C_\Sigma$. Hence, not more than two closely located
peaks from the set (\ref{Vin}) can be well pronounced on the $I-V$ 
curve. In the case $\Delta (T) \agt e^2/C_\Sigma$ the peak position is
close to the center of the subthreshold region, and, hence, close to the 
peak due to the Josephson-plus-quasiparticle  cycle \cite{Fulton-restun}. 

        Peculiarities of another type (jumps of the current) also exist
at $V<V_t$ at finite temperatures. Their positions satisfy equation
$V=2\Delta (T) C_\Sigma /eC_i +V_{i,n}$ 
and correspond to the energy gain $W=2\Delta (T)$ for a particular
tunneling process. The jump height decreases with the decrease of voltage
and vanishes at $V<2\Delta (T) /e$. These jumps of current  
are not well-noticeable in Figs. 1(c) and 1(d).

        Now let us consider the noise-limited sensitivity of the 
SET. The minimum detectable charge for the given bandwidth 
$\Delta f$ is $\delta Q_0=(S_I\Delta f)^{1/2}/(\partial I/\partial Q_0)$ 
where the 
spectral density $S_I$ of the current noise is taken in the low frequency
limit. The ultimate low-temperature ($T\ll e^2/C_\Sigma$) sensitivity in 
the $NININ$ case is \cite{SQUID,Theses} 
$\min\delta Q_{0} \simeq 2.7 C_\Sigma (R_{min}T\Delta f)^{1/2},
\, R_{min}=\min \{ R_1,R_2 \}$. This result can be somewhat improved 
in the $NISIN$ SET (with the same resistances) operating near 
the threshold $V_t$ of quasiparticle
tunneling. At low temperatures, $T \ll \min \{e^2/C_\Sigma , \Delta (T) \}$,
and for $V$ close to nondegenerate $V_t$, we can use approximation 
$S_I \simeq 2eI$, 
$I\simeq I_{0,i} ((V-V_t)C_1C_2/C_iC_\Sigma)$, where  
	$I_{0,i}(v)=(1/eR_i) [T \Delta (T)/2]^{1/2} \int_0^\infty dy/\sqrt{y}/
[1+\exp (y+(\Delta-ev)/T))]^{-1}$
	is the ``seed'' $I-V$ curve of $i$-th junction.
Then the ultimate sensitivity is given by equation
\begin{equation}
\min \delta Q_0 = C_\Sigma (2e\Delta f)^{1/2} \min_{v} \{\sqrt{I_0(v)} /
        (dI_0/dv) \},
\label{dQ0}\end{equation}
        and finally we get the result
\begin{equation}
\min \delta Q_0 \simeq 2.6 C_\Sigma (R_{min}T\Delta f)^{1/2} 
[T/\Delta (T)]^{1/4}
\label{dq0NS}\end{equation}
which is better than $NININ$ sensitivity when $T<\Delta (T)$. The main 
reason for the improvement is the increase 
\cite{Lafarge,Tinkham-SET,Krech-sc} of the transfer coefficient
 $\partial I/\partial Q_0 \simeq (e/C_i)(\partial I/\partial V)$, 
because the differential resistance $R_d$ of the
``seed'' $I-V$ curve near the onset of quasiparticle tunneling
is less than $R_i$.
	Notice that the ``orthodox'' theory used here is valid only
if $R_d \agt R_Q$ because the cotunneling processes \cite{Av-Odin,Krech-sc} 
impose the lower bound for $(\partial I/\partial V)^{-1}$ on the 
order of $R_Q$ \cite{Av-Kor}.
        For relatively high temperatures the ratio of minimum $\delta Q_0$ 
in $NISIN$ and $NININ$ cases is larger than $[\Delta (T)/T]^{1/4}$
(e.g., compare the dashed lines in Fig.\ 2) because $NININ$ sensitivity
starts to deviate up from the low-temperature approximation at smaller 
$T$ than $NISIN$ sensitivity.

        The improvement of the ultimate sensitivity is more 
significant in $SISIS$ SET. For pure BCS model the ``orthodox'' theory
gives infinite derivative $\partial I/ \partial Q_0$
 at $V=V_t$ even for finite temperature leading to $\delta Q_0 
\rightarrow 0$. Hence, the ``orthodox'' ultimate sensitivity depends on 
the imperfection of the current jump which is described in our model
by the energy gap spread $w_0$ ($w_0 \ll \min \{\Delta (T), e^2/C_\Sigma \}$).

        Figure 2 shows $\delta Q_0$ together with  current $I$
and ratio $S_I/2eI$, as functions of the voltage for the symmetric
$SISIS$ SET with parameters $\Delta (0) =0.5 e^2/C_\Sigma$,
$w_0 =0.05\Delta (0)$, $T=0.1e^2/C_\Sigma$, and $Q_0=0.25e$ 
(numerical calculations are done using the method described in Refs. 
\cite{SQUID} and \cite{Theses}). Dashed lines show
$\delta Q_0$ for similar $NININ$ and $NISIN$ SETs. One can
see that the sensitivity of $SISIS$ SET is much better than
for $NININ$ and $NISIN$ cases within a relatively narrow voltage range
which corresponds to the jump of current. 

        In contrast to $NININ$ and $NISIN$ cases, the approximation
$S_I\simeq 2eI$ is not accurate in the vicinity of $V_t$ for $SISIS$
SET even at
low temperatures (see Fig.\ 2) because the relatively large tunneling
rate in the junction determining $V_t$, is comparable to the tunneling 
rate in the other junction. This approximation is valid only if
$T \ll \Delta (T) \ll e^2/C_\Sigma$, and would lead to inaccuracy
typically about 10\% for the analytical calculation of $\min \delta Q_0$
if $T\ll \Delta (T) \sim e^2/C_\Sigma$. Nevertheless, it can be used
as a crude estimate. Using Eq.\ (\ref{dQ0}) and smoothed by $w_0$ 
low-temperature ($T \ll \Delta (T)$) ``seed'' $I-V$ curve for SIS 
junction \cite{Tinkham} we get
\begin{equation}
\min \delta Q_0 \simeq 1.8 C_\Sigma \left( R_{min} \, \Delta f \,
w_0^2/\Delta (T) \right)^{1/2}.
\label{SISIS}\end{equation}
	Notice  that the numerical factor depends on the particular 
model describing the shape of the current jump.
        Comparing Eq.\ (\ref{SISIS}) with the result for $NININ$ SET,
we see that the temperature $T$ is replaced in $SISIS$ case by 
$w_0^2/\Delta (T)$. Hence, the ultimate sensitivity is better in 
$SISIS$ SET (resistances are the same) with sufficiently narrow width of
the current jump, $w_0 < (T\Delta (T))^{1/2}$.

	In the case of very sharp ``seed'' $I-V$ curve, $w_0 \alt \Delta (T) 
R_Q/R_i$, the slope of the jump of the SET $I-V$ curve is determined by
cotunneling \cite{Av-Odin} and it cannot be sharper than crudely 
$R_Q^{-1}$ \cite{Av-Kor}.
Then $\min \delta Q_0$ is on the order of $C_\Sigma (\Delta f \, \Delta (T)
R_Q^2/R)^{1/2}$ (we assume $\Delta (T) \agt e^2/C_\Sigma$, $R_1=R_2$), and 
the ultimate sensitivity is better than for $NININ$ SET if $T \agt \Delta
(T) (R_Q/R)^2$. 
	The sensitivity of such an ideal $SISIS$ SET is even
better than the ``quantum'' ($T=0$) sensitivity of a symmetric ($R_1=R_2$) 
$NININ$ SET operating at $V_t \sim e/C_\Sigma$ (in that case \cite{SQUID}
$\min \delta Q \sim (\hbar C_\Sigma \, \Delta f)^{1/2}$), if $R/R_Q \agt 
\Delta(T) C_\Sigma/e^2$. However, notice that the quantum-noise-limited
$\min \delta Q$ of a $NININ$ SET can be made arbitrary small using 
small $V_t$ (and large ratio $R/R_Q$) \cite{SQUID} or large ratio 
$R_1/R_2$ \cite{Theses}; hence,
in this sense the use of superconducting electrodes cannot improve further
the ultimate sensitivity.

\vspace{1cm}

        The author thanks D.\ V.\ Averin and K.\ K.\ Likharev
for valuable discussions and critical reading of the manuscript.
The work was supported in part by ONR grant N00014-93-1-0880 and
AFOSR grant 91-0445.

        \begin{figure}
\caption{ $I-V$ curves for (a) $NININ$, (b) $NISIN$ (or $SINIS$), and (c)--(d)
        $SISIS$  SETs for three values of $Q_0$ (0, $e/4$, 
        and $e/4$) and several temperatures $T$.
        The curves for different $T$ are offset vertically 
        for clarity. Notice that the modulation by $Q_0$ survives up to
        higher $T$ in the superconducting transistors.}
\label{circuit}\end{figure}  

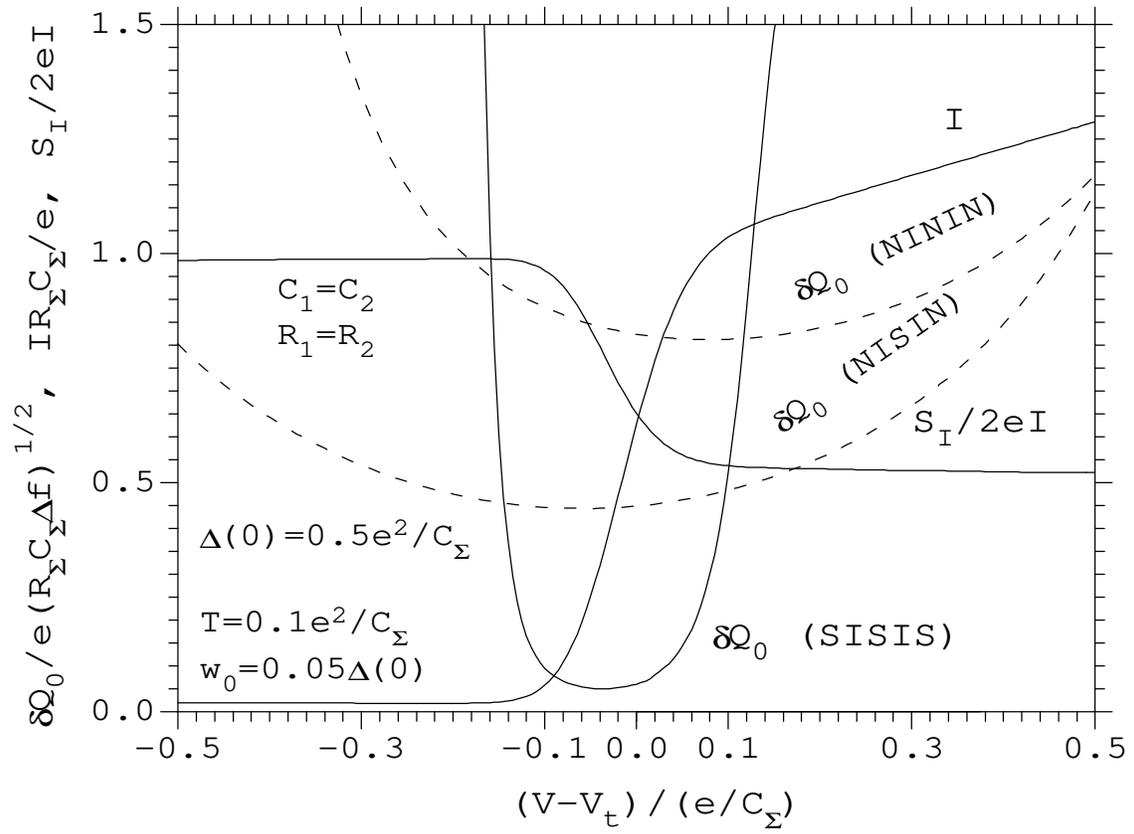
        \begin{figure}
\caption{The minimum detectable charge $\delta Q_0$, the current $I$, 
        and the ratio $S_I/2eI$ as functions of the bias voltage $V$ 
        for $SISIS$ SET. Dashed
        lines show $\delta Q_0$ for $NININ$ and $NISIN$ SETs.
        The best sensitivity is achieved in $SISIS$ case.}

\label{line} \end{figure}


\begin{references}

\bibitem{Av-Likh}  D.  V.  Averin  and  K.  K.  Likharev,  in   {\it 
Mesoscopic Phenomena in Solids}, edited by B. L. Altshuler, P. A. 
Lee, and R. A. Webb (Elsevier, Amsterdam, 1991), p. 173.

\bibitem{Likh-87} K. K. Likharev, IEEE Trans. Magn. {\bf 23}, 1142 (1987).

\bibitem{Lafarge} P. Lafarge, H. Pothier, E. R. Williams, C. Urbina,
and M. H. Devoret, Z. Phys. B {\bf 85}, 327 (1991).

\bibitem{Tinkham-SET} J. M. Hergenrother, M. T. Tuominen, T. S. Tighe,
and M. Tinkham, IEEE Trans. Appl. Supercond. {\bf 3}, 1980 (1993).

\bibitem{Mooij} E. N. Visscher, S. M. Verbrugh, J. Lindeman, P. Hadley,
        and J. E. Mooij, Appl. Phys. Lett. {\bf 66}, 305 (1995).

\bibitem{Krech-sc} W. Krech and H.-O. M\"uller, Mod. Phys. Lett. B {\bf 8},
	605 (1994).

\bibitem{SQUID} A. N. Korotkov, D. V. Averin, K. K. Likharev,
and S. A. Vasenko, {\it Proceedings of SQUID'91} (Berlin, 1991); 
in: {\it Single-Electron 
Tunneling and Mesoscopic Devices}, ed. by H. Koch and H. Lubbig 
(Springer-Verlag, Berlin Heidelberg, 1992), p. 45.

\bibitem{Theses} A. N. Korotkov, Ph. D. Theses (Moscow State University,
        1991); see also A. N. Korotkov, Phys. Rev. B {\bf 49}, 10381 (1994).

\bibitem{Birk} H. Birk, M. J. M. de Jong, and C. Sch\"{o}nenberger,
        Phys. Rev. Lett. {\bf 75}, 1610 (1995).

\bibitem{Hershfield} S. Hershfield, J. H. Davies, P. Hyldgaard, 
C. J. Stanton, and J. W. Wilkins, Phys. Rev. B {\bf 47}, 1967 (1993).

\bibitem{Chen} L. Y. Chen, Mod. Phys. Lett. B {\bf 7}, 1677 (1993);
        {\bf 8}, 841 (1994).

\bibitem{Galperin} U. Hanke, Yu. M. Galperin, K. A. Chao, and N. Zou,
Phys. Rev. B {\bf 48}, 17209 (1993).

\bibitem{Krech} W. Krech and H.-O. M\"{u}ller, Z. Phys. B {\bf 91}, 423
        (1993).

\bibitem{Galperin-SC} U. Hanke, Yu. Galperin, K. A. Chao, M. Gisselfalt,
M. Jonson, and R. I. Shekhter, Phys. Rev. B {\bf 51}, 9084 (1995).

\bibitem{Tinkham} M. Tinkham, {\it Introduction to superconductivity},
        (McGraw-Hill, New York, 1996).

\bibitem{RC-SET} A. N. Korotkov, Phys. Rev. B {\bf 49}, 16518 (1994).

\bibitem{Fulton-restun} T. A. Fulton, P. L. Grammel, D. J. Bishop, and
	L. N. Dunkleberger, Phys. Rev. Lett. {\bf 63}, 1307 (1989).

\bibitem{Av-Odin} D. V. Averin and A. A. Odintsov, Phys. Lett. A {\bf 140},
	251 (1989). 

\bibitem{Av-Kor} D. V. Averin and A. N. Korotkov, in preparation.



\end{references}
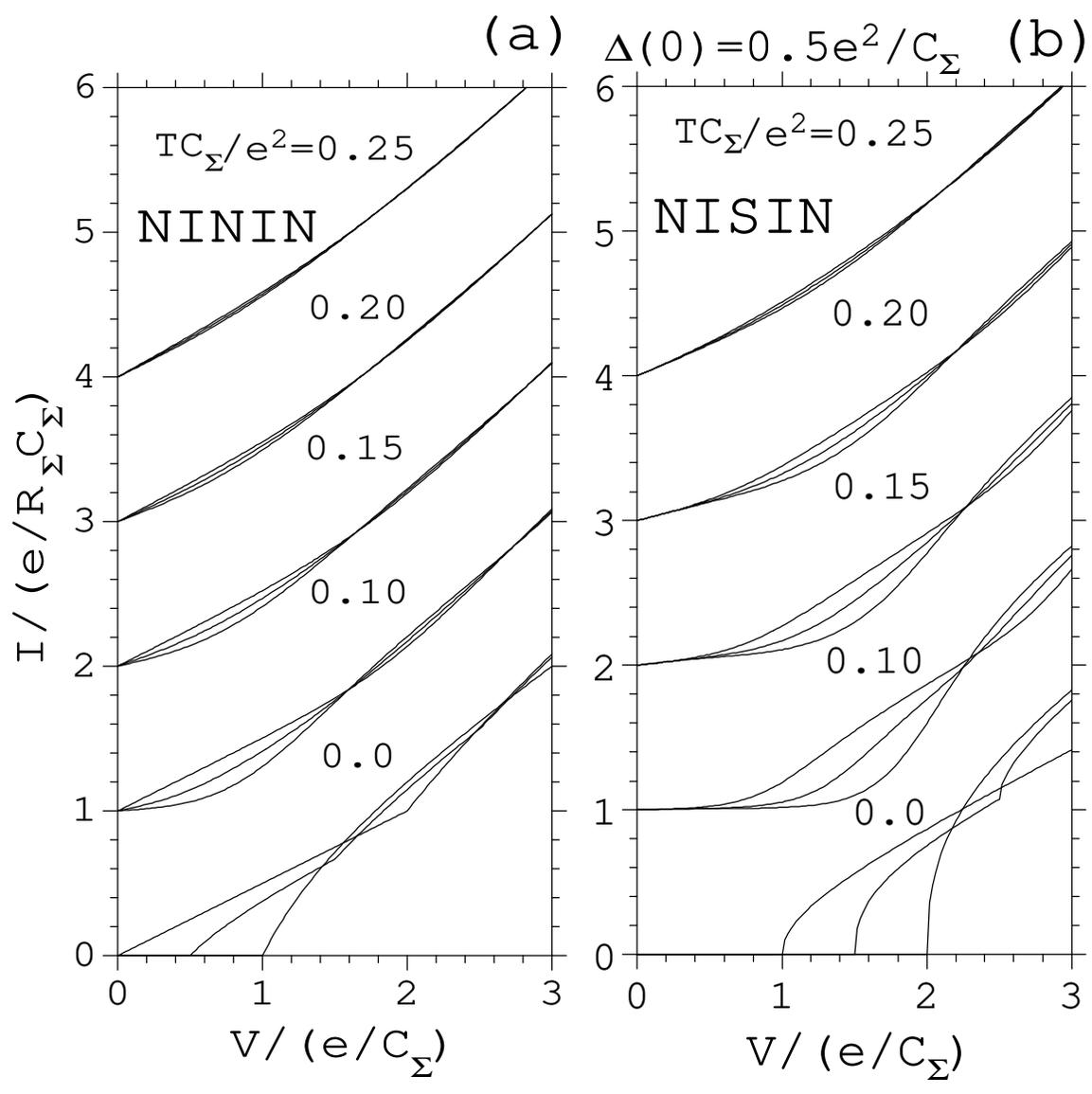
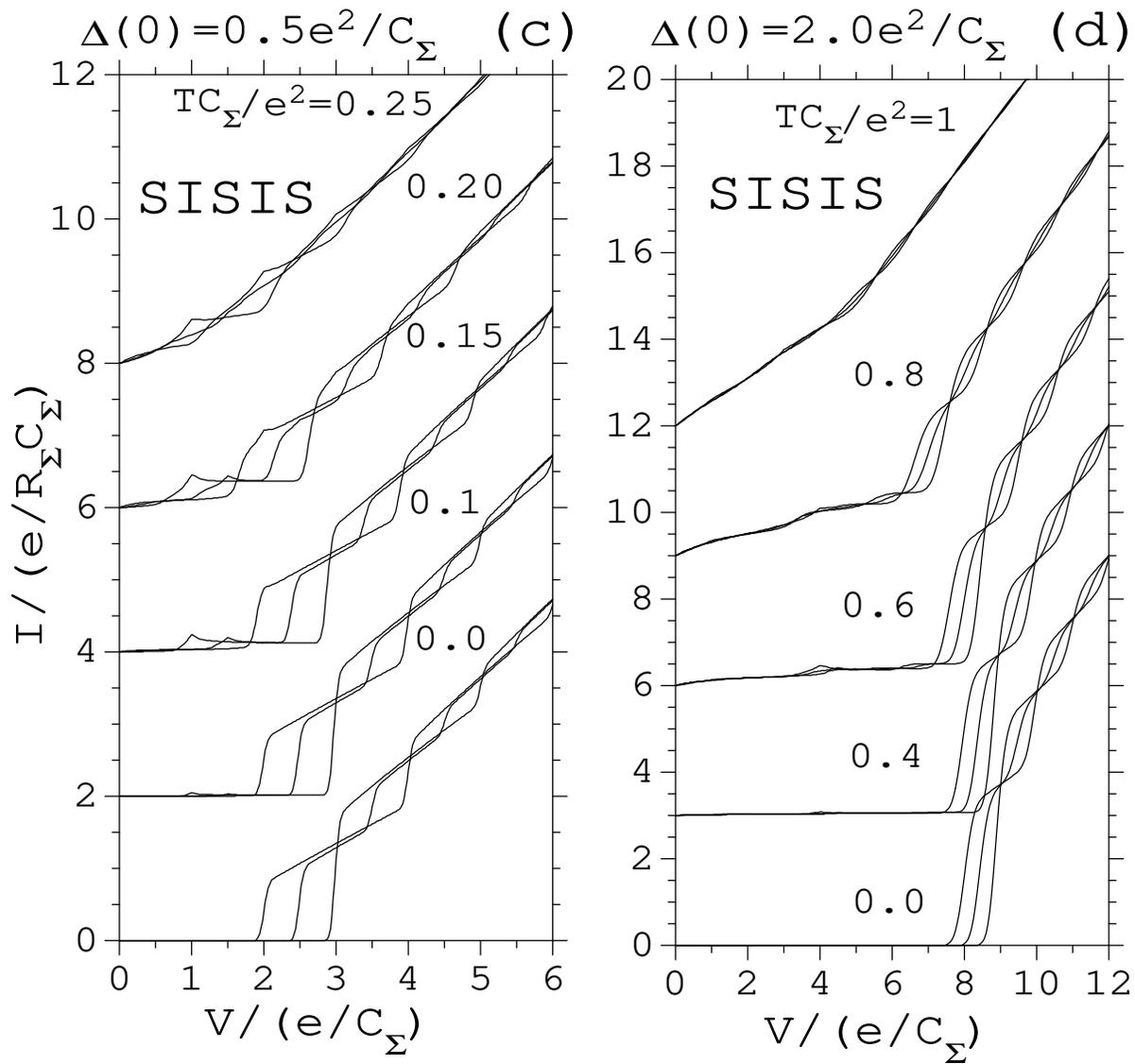
\end{document}